\documentclass[sigconf]{acmart}
\pdfoutput=1
\usepackage[utf8]{inputenc}
\usepackage{array}
\usepackage[
    type={CC},
    modifier={by},
    version={4.0},
]{doclicense}
\usepackage{multicol}
\AtBeginDocument{%
  \providecommand\BibTeX{{%
    \normalfont B\kern-0.5em{\scshape i\kern-0.25em b}\kern-0.8em\TeX}}}

\copyrightyear{2024}
\acmYear{2024}
\setcopyright{rightsretained}
\acmConference[FAccT '24]{The 2024 ACM Conference on Fairness, Accountability, and Transparency}{June 3--6, 2024}{Rio de Janeiro, Brazil}
\acmBooktitle{The 2024 ACM Conference on Fairness, Accountability, and Transparency (FAccT '24), June 3--6, 2024, Rio de Janeiro, Brazil}\acmDOI{10.1145/3630106.3658939}
\acmISBN{979-8-4007-0450-5/24/06}
\begin{document}
%%
%% The "title" command has an optional parameter,
%% allowing the author to define a "short title" to be used in page headers.
\title{Mapping the individual, social, and biospheric impacts of Foundation Models}

%%
%% The "author" command and its associated commands are used to define
%% the authors and their affiliations.
%% Of note is the shared affiliation of the first two authors, and the
%% "authornote" and "authornotemark" commands
%% used to denote shared contribution to the research.
\author{Andrés Domínguez Hernández}
\authornote{Equal contribution as lead authors}
\email{adominguez@turing.ac.uk}
\orcid{0000-0001-7492-7923}
\affiliation{%
  \institution{The Alan Turing Institute}
  \city{London}
  \country{UK}
  \postcode{NW1 2DB}
}

\author{Shyam Krishna}
\authornotemark[1]
\orcid{0009-0008-4370-4725}
\email{skrishna@turing.ac.uk}
\affiliation{%
  \institution{The Alan Turing Institute}
  \city{London}
  \country{UK}}

\author{Antonella Maia Perini}
\authornotemark[1]
\email{aperini@turing.ac.uk}
\orcid{0000-0002-6526-1784}
\affiliation{%
  \institution{The Alan Turing Institute}
  \city{London}
  \country{UK}
  \postcode{NW1 2DB}}

\author{Michael Katell}
\orcid{0000-0003-2200-6246}
\email{mkatell@turing.ac.uk}
\affiliation{%
  \institution{The Alan Turing Institute}
  \city{London}
  \country{UK}}

\author{SJ Bennett}
\orcid{0000-0002-8520-3194}
\email{sj.bennett@durham.ac.uk}
\affiliation{%
  \institution{University of Durham}
  \city{Durham}
  \country{UK}}

\author{Ann Borda}
\orcid{0000-0003-3884-2978}
\email{aborda@turing.ac.uk}
\affiliation{%
  \institution{The Alan Turing Institute}
  \city{London}
  \country{UK}}

\author{Youmna Hashem}
\orcid{0009-0008-2339-1805}
\email{yhashem@turing.ac.uk}
\affiliation{%
  \institution{The Alan Turing Institute}
  \city{London}
  \country{UK}}

\author{Semeli Hadjiloizou}
\orcid{0009-0008-0178-697X}
\email{shadjiloizou@turing.ac.uk}
\affiliation{%
  \institution{The Alan Turing Institute}
  \city{London}
  \country{UK}}

\author{Sabeehah Mahomed}
\orcid{0000-0002-5726-4432}
\email{smahomed@turing.ac.uk}
\affiliation{%
  \institution{The Alan Turing Institute}
  \city{London}
  \country{UK}}

\author{Smera Jayadeva}
\orcid{0000-0002-2050-9731}
\email{sjayadeva@turing.ac.uk}
\affiliation{%
  \institution{The Alan Turing Institute}
  \city{London}
  \country{UK}}

\author{Mhairi Aitken}
\orcid{0000-0002-4654-9803}
\email{maitken@turing.ac.uk}
\affiliation{%
  \institution{The Alan Turing Institute}
  \city{London}
  \country{UK}}

\author{David Leslie}
\orcid{0000-0001-9369-1653}
\email{d.leslie@qmul.ac.uk}
\affiliation{%
  \institution{Queen Mary University of London}
  \city{London}
  \country{UK}}
\additionalaffiliation{
    \institution{The Alan Turing Institute}
    \city{London}
    \country{UK}
}

%%
%% By default, the full list of authors will be used in the page
%% headers. Often, this list is too long, and will overlap
%% other information printed in the page headers. This command allows
%% the author to define a more concise list
%% of authors' names for this purpose.
\renewcommand{\shortauthors}{Domínguez Hernández, Krishna, Perini et al.}
%%
%% The abstract is a short summary of the work to be presented in the
%% article.
\begin{abstract}
  Responding to the rapid roll-out and large-scale commercialization of foundation models, large language models, and generative AI, an emerging body of work is shedding light on the myriad impacts these technologies are having across society. Such research is expansive, ranging from the production of discriminatory, fake and toxic outputs, and privacy and copyright violations, to the unjust extraction of labor and natural resources. The same has not been the case in some of the most prominent AI governance initiatives in the global north like the UK’s AI Safety Summit and the G7’s Hiroshima process, which have influenced much of the international dialogue around AI governance. Despite the wealth of cautionary tales and evidence of algorithmic harm, there has been an ongoing over-emphasis within the AI governance discourse on technical matters of safety and global catastrophic or existential risks. This narrowed focus has tended to draw attention away from very pressing social and ethical challenges posed by the current brute-force industrialization of AI applications. To address such a visibility gap between real-world consequences and speculative risks, this paper offers a critical framework to account for the social, political, and environmental dimensions of foundation models and generative AI. Drawing on a review of the literature on the harms and risks of foundations models, and insights from critical data studies, science and technology studies, and environmental justice scholarship, we identify 14 categories of risks and harms and map them according to their individual, social, and biospheric impacts. We argue that this novel typology offers an integrative perspective to address the most urgent negative impacts of foundation models and their downstream applications. We conclude with recommendations on how this typology could be used to inform technical and normative interventions to advance responsible AI.
\end{abstract}

%%
%% The code below is generated by the tool at http://dl.acm.org/ccs.cfm.
%% Please copy and paste the code instead of the example below.
%%


%%
%% Keywords. The author(s) should pick words that accurately describe
%% the work being presented. Separate the keywords with commas.

%%
%% This command processes the author and affiliation and title
%% information and builds the first part of the formatted document.

\maketitle

\section{Introduction}
Since the release of ChatGPT at the end of 2022, there has been considerable and sustained interest across policy, academia, and public discourses in the risks posed by artificial intelligence (AI), and particularly generative AI. Though ChatGPT may have generated the most column inches, its underpinning model---GPT 3.5 and, subsequently, GPT 4---was just one example of wider developments in the area of so-called “foundation” or “frontier” models.

Foundation models are AI technologies trained on very large, “broad” datasets that can be applied to a wide range of tasks and purposes \cite{bommasani_opportunities_2022}. Their general, unspecified purpose and widespread deployment leads to concerns about potential unanticipated impacts when used in novel areas. These models are considered to form the ``foundation'' or  base architecture of other systems. For example, a number of new applications such as ChatGPT and the conversational features of the Bing search engine, have been built on top of successive versions of OpenAI’s GPT foundation model, which has been designed for natural language processing tasks. 

While the techniques involved in developing and deploying foundation models are not new, the scale of the data used for training, the coordination of global networks of labor to process these massive amounts of data, the increased architectural complexity, and the development of increasingly powerful processors and other computational resources have together made possible new levels of predictive and generative sophistication. Due to the resources required to produce foundation models, the most prominent ones known to exist are the products of large and well-financed technology companies, such as Anthropic, Cohere, Hugging Face, Meta, OpenAI, Microsoft, and Alphabet (the latter two being substantial funders of other firms).

Responding to the rapid rollout and large-scale commercialization of foundation models, large language models (LLMs), and generative AI, an emerging body of work is exploring the myriad impacts these technologies are having across society. Such research is expansive. Prominent topics range from the production of discriminatory, fake, and toxic outputs and privacy and copyright violations to the unjust extraction of labor and natural resources. Foundation models present complex issues for law, policy, and practice, which arise concretely from the intertwined array of socio-technical systems in which their design, development, and use are embedded. These technologies pose risks and hazards that emerge from real-world contexts. Such issues arise from the global supply chains and labor sources that support their production, as well as the market forces and regulatory environments that influence their funding and financing. Risks likewise arise from the sociohistorical realities and legacies of inequity and exclusion that shape the data on which they are trained and from the patterns of privilege and socioeconomic stratification that influence the composition of the project teams that build them. All these as real-world contexts lead to the real-world consequences that have been the subject of much critical and responsible AI research on the impact of foundation models and generative AI over the past year. Yet, despite the importance of centering this empirically oriented work, much of the agenda-setting policy and public discourse emerging primarily from the main geographical centres of innovation on foundation models (and relatedly LLMs and generative AI) has focused on sensational, catastrophic, and largely speculative risks relating to extreme and hypothetical scenarios.

The narrative of catastrophic risks was amplified by much mainstream, English language, media coverage of foundation models throughout 2023, fueled by high profile statements declaring potential existential threats from AI~\cite{metz_godfather_2023, metz_how_2023}. Most notably a statement published in May 2023, and signed by many prominent figures in the field of AI, claimed that “Mitigating the risk of extinction from AI should be a global priority alongside other societal-scale risks such as pandemics and nuclear war” \cite{center_for_ai_safety_statement_nodate}. While the research behind these claims is scant and broadly disputed, this narrative has reached far into predominant policy and regulatory discussions, including high profile international initiatives in the global north like the UK’s AI Safety Summit~\cite{govuk_bletchley_2023} and the G7’s Hiroshima AI Process~\cite{oecd_g7_2023}. Such processes, while presenting valuable opportunities to bring together the international community in discussing the extant risks that AI presents to people, society, and the planet, and corresponding approaches to needed regulation, have been dominated by technical discourses with narrow framings of “safety” and have prioritized discussions of speculative global catastrophic or existential risks of AI. 

Concerns about this narrowing of AI policy and regulatory discourse have been echoed by various academic communities who have questioned the lack of ``ideological and demographic diversity'' of the AI safety field \cite{lazar_ai_2023}; and the privileging of speculative narratives as effective distractions from existing and well-documented risks and harms of AI systems today~\cite{vallor_shrinking_2023, helfrich_harms_2024, gebru_tescreal_2024}. A wealth of evidence demonstrates in particular the disproportionate negative impacts that AI systems have on marginalized communities---communities who are also underrepresented in decision-making processes regarding the design, development, deployment, or governance of AI. These issues, which are linked to historical and entrenched power asymmetries, are now being exacerbated by AI technologies, and risk becoming increasingly more urgent.  

Therefore, through a review of the literature on the harms and risks of foundation models and generative AI, in this paper, we aim to address this visibility gap between real-world consequences and speculative risks. We propose a critical framework capable of adequately illuminating socio-technical risk contexts and attending to the social, political, and environmental dimensions of foundation models and generative AI. Through reviewing the existing evidence base our research maps a diverse set of risks and harms impacting individuals, communities, society, and the biosphere. 

The paper starts with a discussion on the theoretical framework employed to investigate risks and harms in the context of foundation models. This framework abductively informs the methodological approach, which is detailed in section 3. Section 4 present our findings by delineating three distinct levels of algorithmic impacts---individual, social and biospheric. We conclude discussing how the integrative perspective we offer can aid in understanding the negative impacts of foundation models, and can help shape responsible AI futures.

\section{Theoretical Lens: Expanding views on algorithmic risks and harms}
There is a growing and multidisciplinary body of literature which explores the potential risks and harms posed by the relatively recent widespread popularization of foundation models. Although these risks and harms share many similarities with those presented by other types of AI, and indeed other data-intensive technologies more generally, it is crucial to examine how the proliferation and widespread adoption of foundation models may introduce new challenges, as well as further complicate existing ones.

The concept of risk is pervasive in modern societies and has become a central aspect of the AI governance discourse. Sociological studies have argued that theories of risk management and securitization emerged as a means to grapple with the uncertainties of modernity \cite{beck_risk_1992} and “making the future secure and certain” \cite{schuilenburg_securitization_2017}. 
The understanding of risk and harm, as well as how they interrelate, varies greatly across domains.
For legal, policy, organizational, or actuarial purposes, risk is typically approached through quantifiable measures and anticipatory models of potential adverse or harmful events such pandemics, financial losses, health and safety incidents, or operational disruptions~\cite{dionne_risk_2013, power_risk_2009, nist_artificial_2023, mantymaki_defining_2022}. Risk is therefore not inherently an issue of morality and can be understood as a context-dependent and relational concept---that is, expressed in relation to how factors like technical developments, organizational structures, or innovation may adversely impact a desired future and be variously interpreted by different actors~\cite{von_scheve_risk_2023}. For example, consider a company which uses a job applicant screening software and has discovered that it makes racially biased screening recommendations. Where policy makers and civil society may be concerned about the various ways in which the technology undermines individual rights and social equity, a company might instead be more likely to focus on the risk of reputational damage and how this could in turn result in public distrust, low technology adoption, and commercial impacts. 

Critical social science perspectives underline the multiple ramifications of identifying, analyzing, and managing risk associated with the impacts of technology in society~\cite{stahl_organisational_2022, wirtz_governance_2022}. In this, some areas of research critically examine the various ways in which the risks and harms resulting from data and algorithmic systems are patterned across society at different levels \cite{lupton_risk_2013, canning_social_2021}. For instance, the field of science and technology studies (STS) has a well-established body of work exploring the interdependencies between technology and society, challenging the widespread deterministic view of technology as operationally autonomous and inherently value-neutral, and of innovation as following a predictable linear trajectory \cite{mackenzie_social_1999}. When it comes to examining the impacts of AI, scholars in the fields of STS, critical data studies, and socio-legal studies have proposed to conceptualize the harmful effects of data-intensive technologies as issues relevant to human rights, social justice, cultural institutions, the rule of law, the public sphere, economic systems, and the environment \cite{malik_social_2022, crawford_atlas_2021, williams_social_1996,marjanovic_theorising_2022, stoilova_investigating_2021, dominguez_hernandez_co-creating_2023, smuha_beyond_2021}. It is also from a critical perspective that some scholars increasingly highlight the need to approach algorithmic governance not only in terms of speculative notions of risk, but also in response to actual and observed harms to humans and the environment \cite{ferri_risk_2023}. 
%As suggested by Stoilova et al., risk in and of itself is not a necessary or sufficient cause of harm, given that harm can occur due to unanticipated factors and circumstances \cite{stoilova_investigating_2021}; not to mention the difficulties of observing and measuring harm \cite{dominguez_hernandez_co-creating_2023}.

To build an understanding of risks and harms of technology from a moral and ethical standpoint, it is key to consider the questions of how power is distributed across the entire ecosystem of actors, where and how power manifests, and how asymmetries of power and information relate to differential impacts in society. Here it is helpful to draw insights from feminist and data justice scholarship, socio-legal studies, and the environmental justice literature. In particular, the concept of intersectionality makes explicit how social hierarchies manifest in different outcomes and experiences depending not only on the identity of the impacted individual or group, but also on the multiple compounding intersections of social identity and standing. These intersections go beyond crude constructions and classifications of individuals in society, to recognize the ways in which nuances of context, identity, and circumstance distinctly shape people’s understanding and experiences of harm \cite{crenshaw_mapping_1991}. 

To capture these differential impacts as they relate to the proliferation of foundation models, in this paper we propose, following Leslie and Rossi \cite{leslie_acm_2023}, to assess risks and harms according to their individual, social, and biospheric levels of impact, contributing to an emerging dialogue within the FAccT and AI ethics communities \cite{simbeck_facct-check_2022, weidinger_taxonomy_2022}. This perspective allows for a nuanced understanding of the observed and anticipated impacts of foundation models by recognizing the inherent socio-technical and globally entangled nature of these technologies. We use this framework to map the risks and harms found in the literature to date and offer an integrative perspective to address the most urgent negative impacts of foundation models and their downstream applications.

\section{Methods: snowball and structured search}

We conducted a review of literature using two complementary and reinforcing qualitative literature review approaches which have been shown to be useful in studies aimed at surfacing patterns and trends in corpora of academic research~\cite{wohlin_successful_2022}. 

We first applied a snowball approach \cite{wohlin_guidelines_2014}, crowd-sourcing among all the co-authors, a core set of recent papers focused broadly on algorithmic harms and risks \cite{bender_dangers_2021, birhane_science_2023, bommasani_opportunities_2022, derczynski_assessing_2023, jones_capturing_2022, kasirzadeh_conversation_2023, shelby_sociotechnical_2023, weidinger_taxonomy_2022}. These papers were selected based on our inclusion criteria of covering a broad range of multidisciplinary literature addressing risks and harms associated with foundation models. Populating a Zotero database, we followed snowball sampling approaches starting from the set of core articles and then carrying out a manual search using the references in those articles. Forward snowballing was used to identify new papers that focus on algorithmic risks and harms citing those in the core list using Google Scholar and Internet-based searches to access the abstract and full-text. We identified a total of 64 articles including journal articles, conference papers, and pre-prints drawing broadly on technical and socio-technical literature in which key risks and issues of foundation models, LLMs, and generative AI are described historically and contemporaneously. The search inclusion criteria covered English language publications and the most recent publication was of October 2023. 

Complementing the snowball approach, we conducted a structured database search to identify primarily academic articles with a focus on the risks and harms associated with foundation models, LLMs, and generative AI. 

Inclusion criteria covered English language journal articles, including conference papers and pre-prints, but excluding grey literature, monographs, commentaries, correspondences, and opinion pieces. The following electronic databases were searched: arXiv, ACM Digital Library, IEEE, Scopus, and Web of Science.\footnote{The ACL Anthology database was not used as a source for the search given this database's core focus on computational linguistics and natural language processing.} 
After the removal of duplicates and non-relevant papers, there were 114 papers in total for analysis, of which 11 were also papers in the snowball search results. The resulting set of papers were published in the period January 2018 to July 2023. The start date aligns with the date of appearance in the literature of transformer architectures which underpin LLMs, such as the BERT large language representation model \cite{vaswani_attention_2017}.

We applied an abductive approach \cite{tavory_abductive_2014, vila-henninger_abductive_2022} to analyze the combined corpus of literature 167 papers. The abductive approach was used to code the results and subsequently map key relationships between the risks and harms found in the literature guided by our theoretical framework described in Section 2. Firstly, we extracted applicable keyword information from the abstract of each primary study as part of an initial coding process. The resulting keywords served largely as summative and process attributes of a risk or type of harm relevant to foundation models or LLMs. Three researchers subsequently clustered these attributes and assigned them into parent categories. The codes were subsequently refined by all co-authors (see Appendix for search strategy). In this stage, we considered whether at least one of the risks and harms listed in the paper implied or directly referred to individual, social, or biospheric impacts to map the papers onto our three-level framework. These codified findings are further analyzed in Section 4.

\subsection{Limitations}

We note that our structured search showed a prevalence of preprint papers (non-peer review) in the result set, some of which had been cited extensively in both preprint and peer-review articles.\footnote{We made a distinction between citations of preprints referenced by other preprints, some of which exceeded 50 citations, and those referenced by peer-reviewed publications. Those preprints that were more extensively cited had citations in both preprints and peer-reviewed articles.}
We used citations only as a broad indication of the preprints' impact. While some preprints were highly cited, these may have not undergone a rigorous peer-review process, and their findings should be interpreted with caution. In our sample, we analyze articles solely as an indicator of the range of concerns raised in the existing literature. The selected papers were also limited to the English language, and date range of January 2018 to July 2023. This range by default provided only a snapshot of a growing and multidisciplinary body of work which is arising in response to the fast development and roll-out of foundations models and LLMs, especially generative AI related studies. In addition, our structured literature search, which is not intended as systematic, only centred academic articles as a relevant account of the discourse on risks and harms of foundation models and LLMs to inform governance. We also note that this body of research is not focused exclusively on risks and harms, but often addresses, or points to, mitigations and governance of foundation models and AI at large. An analysis of other non-academic and informal sources was not part of this study and it warrants additional research. Notwithstanding, the codified mappings provide a substantive view of the research landscape and the range of concerns that are being raised around foundation models and LLMs, their development, implementation, and applications. 

\section{Mapping individual, social, and biospheric impacts of foundation models}

The result set of 14 risks and harms coding (Table \ref{tab:mapping}) is telling of the wide-ranging and multifaceted challenges posed by foundation models and AI systems at large. Papers typically engage with three or more aspects, exemplifying the broad spectrum of concerns in the literature to date. For instance, Zhuo et al.  \cite{zhuo_red_2023} delve into potential biases, toxicity, and issues related to the reliability and robustness of ChatGPT. Notably, a substantial set of search results focuses on bias and discrimination. These are most typically related to LLMs, and include works such as van der Wal et al. \cite{van_der_wal_undesirable_2023} and Huang et al. \cite{huang_trustgpt_2023}, which address multiple biases and harmful stereotypes, and others, like Abid et al. \cite{abid_persistent_2021}, Felkner et al. \cite{felkner_winoqueer_2023}, Ovalle et al. \cite{ovalle_im_2023}, and Gadiraju et al. \cite{gadiraju_i_2023} that concentrate on specific societal biases. The result set also reveals that papers that focus on specific or a limited number of risks and harms often scrutinize issues of unreliable performance of foundation models, or misinformation and propaganda. While other issues such as lock-in and opacity, or overdependence in human-computer interaction, are less prevalent in the result set, it is important to note that their absence does not imply that they are not identified as relevant within the existing literature. Instead, they are simply not the primary focus of most papers in our dataset. 

In the following, we elucidate three distinct levels of algorithmic impact (individual, social, and biospheric) with examples stemming from the literature and categorize the risks and harms emerging from our search using this typology (see Table \ref{tab:mapping}). This lens on algorithmic harm aims to account for the intersectional dimensions and socio-technical embeddedness of foundation models with the aim to better inform potential technical and normative interventions to advance responsible AI.

\subsection{Individual risks and harms}

It is now undeniable that AI systems can have adverse impacts to individuals which may be physical, psychological, or financial, but may also negatively affect a person’s dignity, reputation, or fundamental rights and freedoms. Indeed, individual-level risks and harms arising from the implementation of data-intensive technologies have been largely documented in the literature, with individual impacts having immediate as well as long-lasting implications for those at the receiving end of such harms \cite{raymond_beyond_2017}. Equally so, a focus on individual rights has been at the core of conceptions of data protection, privacy, and security within digital regulation such as the European Union’s General Data Protection Regulation (GDPR), where protection and redress mechanisms privilege notions of personal data or data subjects \cite{smuha_beyond_2021}.

%insert table here

\begin{table}
  \caption{A mapping of foundation models risks and harms according to their level of impact (individual, social, biospheric). The table shows the number of papers across categories of harm and levels of impact, and the percentage (in brackets) of the  total number of papers in the set in which the risk/harm appears. Papers appearing in more than one category were counted more than once. See Appendix for an explanation of each risk and harm category.}
  \label{tab:mapping}
 \small % Smaller font size
\begin{tabular}{p{0.45\linewidth}  p{0.1\linewidth}  p{0.1\linewidth} p{0.1\linewidth}}
\hline \\[-1.8ex] 
\textbf{Risks and Harms Category}& \textbf{Individual} & \textbf{Social} & \textbf{Biospheric}\\ 
\hline \\[-1.8ex] 
Bias and societal prejudices & 71 (42.5\%) & 71 (42.5\%) &0 \newline(0\%) \\ \hline
Misinformation, disinformation and propaganda & 45 (26.9\%) & 45 (26.9\%) & 1 \newline (0.6\%) \\ \hline
Unreliable performance & 43 (25.7\%) & 40 (24\%) & 0 \newline (0\%)\\ \hline
Cybersecurity risks and harms  & 37 (22.2\%)& 27 (16.2\%)& 0 \newline (0\%)\\ \hline
Privacy risks and harms & 28 (16.8\%) & 21 (12.6\%) & 0 \newline (0\%) \\ \hline
Systemic social and economic risks and harms & 18 (10.8\%) & 24 (14.4\%) & 1 \newline (0.6\%) \\ \hline
Legal and regulatory violations & 17 (10.2\%) & 17 (10.2\%) & 0 \newline (0\%) \\ \hline
Environmental effects and ecological disruption & 8 \newline (4.8\%) & 8 \newline (4.8\%) & 19 (11.4\%) \\ \hline
Misuse
& 13 (7.8\%) & 14 (8.4\%) & 19 (11.4\%) \\ \hline
Lock-in and opacity risks
& 13 (7.8\%) & 15 \newline (9\%) & 0 \newline (0\%) \\ \hline
Overdependency in human-computer interaction  & 11 (6.6\%) & 6 \newline (3.6\%) & 0 \newline (0\%) \\  \hline
Data risks and harms
& 7 \newline (4.2\%) & 7 \newline (4.2\%) & 0 \newline (0\%) \\  \hline
Value misalignment
& 2 \newline (1.2\%) & 2 \newline (1.2\%) & 0 \newline (0\%) \\ \hline
Extreme or catastrophic risks and harms 
& 2 \newline (1.2\%) & 2 \newline (1.2\%) & 1 \newline (0.6\%) \\ 
\hline \\[-1.8ex] 
%\multicolumn{4}{p{20pc}}{*The papers under this category refer to speculative far-reaching or irreversible adverse impacts of foundation models at societal scale
%that extend beyond immediate impacts.}\\
\end{tabular} 
\end{table} 

One of the biggest ethical concerns associated with AI systems has been their role in creating discriminatory, biased, and unfair outcomes which impact an individual’s safety, health, wellbeing, and integrity. The proliferation of applications based on foundation models has not only augmented these concerns but has given rise to a host of new ethical, privacy, and safety issues. Biases and social prejudices present in training datasets and encoded in foundation models trickle down to applications like chatbots or image generators, thereby reproducing existing patterns of harm, exclusion, and discrimination \cite{abid_persistent_2021, steed_upstream_2022, gadiraju_i_2023}. For instance, chatbots---such as OpenAI’s ChatGPT or Google’s Bard---are capable of producing natural language responses to prompts through complex statistical associations \cite{bommasani_opportunities_2022, zhao_survey_2023}. These responses are enabled by the vast amounts of data that the foundation models have been trained on, comprising mostly of internet-scraped datasets that have been shown to encode biased and harmful associations across protected categories of religion, disability, gender, race, and ethnicity \cite{nadeem_stereoset_2020, abid_persistent_2021, deshpande_toxicity_2023}, and also at the intersections of these characteristics \cite{tan_assessing_2019, guo_detecting_2021}. 

When it comes to issues of privacy and online safety, LLMs have been found to be effective tools for fraudulent activities such as scams, impersonation, and phishing, given their capability to generate personalized and highly convincing messages at scale \cite{hazell_large_2023, derner_beyond_2023, neupane_impacts_2023}. Research has also shown that LLMs are prone to leaking private or sensitive information contained within their training data corpus \cite{carlini_quantifying_2023, kandpal_deduplicating_2022}. Because these models are trained on data scraped from the Internet, privacy violations might take place unbeknownst to impacted individuals thereby limiting their ability to consent or opt out \cite{oneill_amplifying_2023}. Furthermore, models can associate different pieces of data to an individual, which poses privacy and security risks wherein sensitive information or even wrongful associations are revealed without consent \cite{huang_survey_2023}. Indeed, in some instances, chatbots have been used with the intention of extracting protected information and bypassing restrictions through the use of malicious methods like prompt hacking, jailbreaking \cite{rao_tricking_2023, wang_decodingtrust_2024}, and prompt injection \cite{perez_ignore_2022, greshake_more_2023}. 

Lastly, the proliferation of foundation models complicates the already significant negative impacts digital technologies have on people’s psychological wellbeing. In the last decade, researchers have raised concerns over how hyperconnectivity and unhealthy relationships with technology can have serious mental health implications and influence how people form individual identities and construct self-hood \cite{albrechtslund_spaces_2013, lupton_quantified_2016, brubaker_digital_2020}. These issues are now exacerbated with the emergence of personalized AI chatbots capable of producing human-like responses. New research into the use of these conversational agents for mental health support warns about the potential for people’s overdependence on these agents and the risk of social isolation \cite{ma_understanding_2023}. Similarly, documented cases of users self-harming as a result of exposure to toxic interactions with AI agents evidence the devastating impacts these technologies can have on vulnerable individuals \cite{walker_belgian_2023, xiang_he_2023, lee_speculating_2023}. For example, prior research has demonstrated that individuals who are isolated may be more vulnerable to misleading information on social media platforms \cite{bonsaksen_loneliness_2021}. When individuals are subject to algorithmic nudging and personalized information campaigns, they are at risk of loss of autonomy, dignity, and integrity \cite{rosenberg_metaverse_2023}.

\subsection{Social risks and harms}

While the rise of foundation models has generated huge expectations about their positive transformational potential for society in many domains, there are also concerns about how their adoption can contribute to worsening existing social inequalities, biases, and injustices. Beyond individual risks and harms, there are different ways in which technology can have wider collective and societal impacts. One way to conceptualize collective risks and harms is to look at the compounded or aggregate impacts to a particular group, or to the functioning of a society as a whole \cite{malik_social_2022}. 

Risks and harms to communities can be characterized by how particular demographics and group formations within society are distinctively impacted by technology, particularly when it comes to vulnerable groups with protected characteristics, e.g., migrants, refugees, LGBTQI+ communities, Muslims, the elderly. \footnote{We note that notions of community are complex, often comprising loosely held webs of relational experience and knowledge, which share geo-spatial, identity-based, and/or digital spaces amongst other facets, with community membership subjective and fluid \cite{bryer_promoting_2020}. Here, communities are defined as groups sharing characteristics pertaining to intersectionality, power, and positionality \cite{ruhland_positionality_2023}.} As we have shown, not only can these technologies produce responses that are discriminatory and psychologically triggering for individual users, but the reproduction of hegemonic views, harmful stereotypes, and biases can have detrimental consequences at the collective level, affecting marginalized groups and communities including those who are not direct users, or are far removed from the technology. For instance, Abid et al. \cite{abid_persistent_2021} demonstrate the persistent religious bias present in foundation models, noting that prompts containing the word ``Muslim'' produced responses with violent language. These negative associations, in turn, can spur targeted hate and disinformation campaigns centered around anti-Muslim or Islamophobic sentiments.

Collective forms of harm can also be understood from a relational lens. The relational lens captures the adverse impacts to social relationships of interdependence, trust, and solidarity, thereby negatively impacting people’s experiences of belonging, or their capacity to flourish through their relationship to others. A rich and interdisciplinary body of research has shed light on the range of interpersonal effects arising from people’s interactions with algorithms, including the affective impacts of the algorithmic sorting of content on social media feeds \cite{eslami_i_2015, bucher_algorithmic_2017, stark_algorithmic_2018} and the proliferation of misinformation as detrimental to the quality of the public sphere and democratic deliberation \cite{bennett_disinformation_2018, tsfati_causes_2020, vaccari_deepfakes_2020}. Foundation models in particular can increase the scale and speed at which disinformation campaigns can be disseminated across the information ecosystem \cite{solaiman_release_2019, pan_risk_2023}. As generative AI applications powered by foundation models flood the public sphere with fake information, there is a risk of eroding public trust in the information that circulates online, further fueling social polarization and the creation of echo chambers \cite{kirk_personalisation_2023, simmons_moral_2023}. 

Lastly, algorithmic systems are inextricably embedded in wider social, political, and institutional structures, and as such they also have a bearing on those structures \cite{williams_social_1996}. To grapple with the impacts that foundation models can have at a structural level, one must consider geopolitical, socioeconomic, legal, and organizational manifestations of power \cite{leslie_acm_2023}. For instance, some scholars have warned about the tight concentration of power arising from the limited pool of actors controlling technological capabilities globally \cite{crampton_digital_2019, deibert_mutual_2019, ohara_four_2021}; the influence of these power asymmetries on the development of international policies, standards, and regulations for technological practices \cite{baik_data_2020, chomanski_missing_2021, cohen_between_2019}; and the overwhelmingly invasive data practices carried out by governments and companies \cite{eubanks_automating_2018, fourcade_learning_2020, treguer_seeing_2019}. 

The collective and deeper structural effects of these technologies are difficult to anticipate and measure as they manifest over time as AI becomes more widespread and embedded in society. Nonetheless, it is helpful to analyze and situate foundation models in the context of well-known social implications linked with digital technologies. For instance, going back to the issues related to the public sphere, one could argue that the effortless production and spread of targeted misinformation and propaganda enabled by foundation models, can have lasting effects on wider cultural, social, and political structures, namely established democratic processes, representation, participation in public life, and institutional trust. Given the evidence of public and private actors deploying technologies to generate manipulative content to sway public opinion \cite{funk_freedom_2023, collier_influence_2022}, scholars have warned about an increasing risk to social cohesion given the mounting erosion of trust in democratic structures \cite{kreps_how_2023}.

\subsection{Biospheric risks and harms}

Lastly, biospheric risks and harms refer to the potential adverse impacts to ecological systems and environments, as well as wider impacts to the biosphere at the planetary level \cite{potter_fragmented_1999}. Importantly, this category also includes harms to the individuals and communities negatively impacted by environmental effects and ecological disruptions resulting from AI development and use. In other words, biospheric, social, and individual risks and harms are fundamentally intertwined. A good entry point to understand biospheric impacts is to look at the largely hidden socio-material entanglements of data-intensive systems. That is, the complex supply chains, labor, and underpinning materials crucial to AI infrastructures---including the Rare Earth Elements (REEs) that are necessary to manufacture the semiconductors essential to computer hardware. The global nature of the supply chains of foundation models pose individual, social, and environmental risks by entangling processes of extraction, knowledge production, and ultimately decision-making \cite{rella_close_2023}. 

The development of supercomputers which utilize REEs, and enable foundation models to be trained, has been found to have significant material footprints globally \cite{anton_superconducting_2020}. The mining of silicon and other rare metals not only cause environmental harms, but can rupture local communities \cite{crawford_atlas_2021, al_rawashdeh_socio-economic_2016} and cause individual harm to the workers mining these elements \cite{white_exposure_2016}. Furthermore, the negative consequences are usually borne disproportionately on vulnerable or marginalized communities largely located in the global south \cite{bullard_threat_1993}. Thus, when put in the context of global infrastructures, the interrelated nature of these risks and harms becomes clearer. Robbins and van Wynsberghe argue that the “interconnectedness of AI with ecological, social, and economic systems” \cite{robbins_our_2022} pose very real risks regarding societal lock-in to AI infrastructures. That is to say that beyond a certain point, the nature of the supply chains and materials powering certain AI infrastructures cannot be changed, whilst ecological harms would continue to multiply because of resultant carbon emissions, REE mining, and other processes. Given these potential path dependencies, we are currently at a crucial juncture for developing ethical governance mechanisms that address the interlocking risks of the biospheric and human facets of AI ecosystems.

The development and implementation of data-intensive systems like foundation models is also known to have high local environmental costs in the form of energy consumption required to train increasingly large and complex machine learning systems \cite{kirk_personalisation_2023, luccioni_estimating_2022}, and heavy water consumption needed for data center cooling \cite{li_making_2023}. The amount of compute needed to train large-scale models has doubled every 3.4 months since 2012 \cite{amodei_ai_2018, hao_computing_2019}, which translates to higher energy expenditures, and the use of costly resources even if the improvements in model accuracy are modest (i.e., diminishing returns in terms of model accuracy come at increased computational cost, see \cite{schwartz_green_2019}). For example, the carbon emissions of training Google’s BERT were roughly those of a transatlantic flight \cite{strubell_energy_2019}. All of this at a time when curbing our global emissions is crucial to slowing climate change down and effectively mitigating its effects \cite{united_nations_goal_nodate}. Efforts to develop responsible technologies that minimize their cascading impacts on the environment are hindered by the difficulty in comprehensively assessing carbon and energy footprints of training large models. Tools have emerged enabling practitioners and developers of these energy intensive technologies to monitor and track their models’ emissions \cite{anthony_carbontracker_2020}.

Although the environmental costs of AI are borne by everyone on the planet in terms of negative externalities contributing to climate change and ecological deterioration, they are not uniformly distributed among the world’s populations or regions. The allocations of benefits and risks replicate existing patterns of environmental injustice, coloniality, and ``slow violence'' \cite{nixon_slow_2013}, in which a disproportionate exposure to risks and harms is borne by marginalized communities; be these in terms of pollution, destruction of local ecosystems, or involuntary displacement. By contrast, the majority of the economic benefits reaped by the use of the model mostly go to the model’s proprietary owner(s), even when these technologies adopt an open source ethos \cite{liesenfeld_opening_2023}.

Finally, the impact of pollution and ecosystem degradation on interdependent and connected biological systems (including sentient non-human animals) cannot be ignored \cite{strubell_energy_2019, coghlan_harm_2023}. AI has the potential to put non-humans at further risk (e.g., animals and the ecological systems in which they live) through surveillance, monitoring, and testing for which the benefits are largely directed to humans and society. Potential harms arise from data collection on animals in intensive factory farming to increase productivity, or data tracking of protected wildlife, which hunters and poachers can hack to illegally hunt or trade animals \cite{bossert_animals_2021, coghlan_harm_2023}. Beyond these physical harms to animals, the functional use of AI in image recognition and as recommender systems demonstrate biases in differentiating species borne out from underlying social factors \cite{hagendorff_speciesist_2023}. There is an urgent need to understand and respond in a systemic way to the interconnected and shared vulnerabilities of humans, animals, and the ecosystems in which we live and share \cite{coghlan_harm_2023}.

\subsection{Summary of findings}

Our review shows that a large proportion of papers (40\%) raise concerns about the potential for foundation models to perpetuate and amplify hegemonic views, harmful stereotypes, and societal and behavioral biases on an unprecedented scale. A similarly large set of papers (20\%) also point to issues related to the creation and spread of misinformation and propaganda as well as the potential exploitation of these models for fraudulent services and cybersecurity attacks by malicious actors. Furthermore, the individual and societal impacts stemming from inconsistent or undesirable performance constitute a significant part of the mapped academic literature, making the case for cautious approaches to the deployment and use of such models. 

Although less prevalent, a subset of the literature mapped point to concerns in connection to a range of technical and supply chain challenges of foundation models, and the potential compounding effects of those challenges as foundation models become more widespread. These range from the reliance on proprietary software and lack of transparency that enable lock-in and opacity risks to issues like outcome homogenization, narrowing of the market or monopolization, and the perpetuation of inequalities, which affect societal and economic structures. Our mapping also reveals that approximately 10\% of the mapped papers address the environmental risks and harms associated with foundation models. While the primary focus of these latter works is on the biospheric-level impacts, some authors explore the interconnections between the adverse impacts of foundation models on ecology and animals and their consequential impacts on individuals and societies. This underscores the need for expanded exploration and attention to holistically understand the complexities surrounding the environmental adverse impacts of foundation models. Lastly, despite the extensive emphasis in much of the governance discourse, less attention is given in the academic literature to the extreme scenarios of existential, catastrophic, and other speculative risks of foundation models. These findings highlight the current visibility gap between real-world consequences and speculative risks, and sheds light on the areas requiring urgent and greater attention and efforts.

\section{Discussion: Grappling with the scale and interconnectedness of foundation models}

Foundation models are highly consequential technologies which have sparked discussions about a spectrum of transformational impacts from the most promising to the most concerning. They pose unprecedented governance challenges but also offer an opportunity to draw lessons from, as well as re-examine, the landscape of socio-technical impacts of data-intensive technologies. While emergent and novel, foundation models do not arrive de novo; they are “built on an installed base” \cite{star_ethnography_1999} and inherit the tools and methods of prior generations of machine learning and neural network technologies. As such, many of the foundation-model-related risks and harms that we highlight are likely to appear familiar because they occur, to some degree, in adjacent and precursor technologies. 

Even so, two key differentiating characteristics of foundation models are their massive scale and widespread embeddedness. Foundation models comprise hundreds of billions of parameters, trained on mountains of data, that consume enormous resources for both training and deployment. In particular, the scale of foundation models means that the risks and harms they present are not only likely to be magnified and amplified, but that this will happen in ways which transcend national and political boundaries, requiring a multi-pronged and transnational response. Harms that may have been minimal or just minimally attended to in prior generations of technology are now made visible and urgent. For example, data-intensive systems have been criticized for some time for their carbon footprint \cite{stoll_carbon_2019}, but the high visibility and widely publicized demands of foundation models have brought this issue to the fore to the point that it is unsurprising to see this aspect discussed in the technical literature \cite{bommasani_opportunities_2022}. 

Another differentiating characteristic of foundation models is their embeddedness. Foundation models are conceptualized and architected as the base models for many and diverse types of downstream applications. The embeddedness of foundation models renders them invisible yet pervasive. As a result of their platformized architecture, foundation models form the basis of many thousands of extensions, and as such, the negative impacts and harms stemming from foundation models may be obfuscated and rendered relatively intractable. These two characteristics---scale and embeddedness---position foundation models to be both highly adaptive, highly elusive, and highly dangerous. We argue that any assessment of risks and harms should account for these socio-technical interdependencies, and any design of mitigations and policy responses should be commensurate with the level of impact be it individual, societal, or biospheric. The value of our proposed framework for conceptualizing risks and harms is that by decomposing these into individual, social, and biospheric impacts, we provide a conceptual tool with which to challenge attempts to narrow down salient risks and harms in a way that is meaningful only to some discrete set of affected actors and thereby limited in effect. The prevailing international policy discourse focusing on the technical safety of “frontier AI” systems, which has led to a flight from confrontation of the full spectrum of hazards presented here, is a signal example of such an erroneous narrowing~\cite{helfrich_harms_2024}. In this paper, we underline the necessity of the opposite approach. When identified at their empirical sources, the risks and harms of foundational models become visible from all angles. They cut across individual, collective, and environmental levels, spreading over and affecting entire populations, including entire socio-material and biophysical ecologies of humans and non-humans. In this way, the impact and importance of the full range of risks and harms cannot be invisibilized or waved away in pursuit of economic, geopolitical, or other short-term goals. 

\subsection{Visibility gaps in the current assessment of algorithmic impact}

As we have shown in this paper, the literature on risks and harms of foundation models is expansive and is likely to continue to grow. Grappling with such a vast and heterogenous landscape is a challenging task both for those who are attempting to build a nuanced understanding of the technology as well as for those who are seeking to inform the debate on mitigations and governance. We recognize notable attempts in the literature to taxonomize these risks \cite{weidinger_taxonomy_2022}, as well as significant progress being made internationally to reckon with the scale and embeddedness of foundation models by framing AI governance as a human rights and transnational issue \cite{intelligence_cai_draft_2023,unesco_taking_2023}. However, some of the most prominent governance initiatives, particularly in Europe and the US, has thus far fallen short of contending with the most problematic harms stemming from unequal patterns of data, labor, and resource extraction and instead has focused largely on risks to adoption, technical safety issues, and catastrophic risks.

One of the biggest challenges in assessing and anticipating algorithmic harm has to do with limited evidence and difficulties related to observing the indirect manifestations of harm, as well as foreseeing its effects over time. Grasping these complexities involves nuanced and context-dependent understanding. For instance, there are direct measurable impacts from the energy used to train or operate an LLM which are immediate and visible, but indirect harms arising from AI applications may only come to light over time or when enough evidence and research is made available. An AI system enabling the unsustainable extraction of mineral sites or unfair labor practices in the data labelling supply chain, are just a couple of illustrations of how negative indirect harms are likely to be largely hidden from view \cite{oecd_measuring_2022}. Direct harms to individuals may translate to indirect harms to society and vice versa. For instance, the potential harm that can ensue with data misuse or privacy breach (e.g., personal data) is not just limited to the individual who is directly manipulated, but indirectly affects the interests of society at large \cite{smuha_beyond_2021}. Where anthropocentric and species biases exist in AI, there are further examples of direct harm which remain understudied, largely because wider biospherical needs are not part of the conversation \cite{coghlan_harm_2023}. Algorithmic systems that ignore animals or privileges a particular view of animal welfare while ignoring others can exacerbate these consequential impacts, and not least can lead to indirect and direct harms to interconnected ecosystems shared by both humans and animals \cite{bossert_animals_2021, coghlan_harm_2023}.

\subsection{Building a socio-technical scaffold to technical interventions}

The current strategies for the mitigation of foundation model generated risks and harms that have most traction among policymakers and governments predominantly focus on technical interventions~\cite{oecd_initial_2023}. There is thus an urgent necessity to delve into the social context within which these interventions are situated. This is crucial, especially since the advent of generative AI represents not only a significant milestone in technical advancement but also simultaneously transforms the very fabric of far-reaching socio-digital infrastructure like the Internet and its social experience \cite{heikkila_how_2022, heaven_generative_2022}. 

Consider, for instance, the issue commonly framed in technical terms as “model collapse” or “data pollution”. In this scenario, the generative output from the widespread experimental use of chatbots and image generators feeds into public data pools. This influx poses the risk of influencing future datasets that AI models will subsequently incorporate. The amalgamation of human and AI-generated content potentially undermines the quality and diversity of AI-generated outputs \cite{martinez_towards_2023}. While this cycle undoubtedly necessitates technical intervention, the risk typology outlined in this paper enables the identification of concurrent impacts within the social sphere---impacts like the undermining of the long-term integrity of the information ecosystems on which modern democratic ways of life rest. 

The integrative perspective we offer emphasizes, for example, that the risks surrounding the AI-enabled generation of biased or harmful images go beyond just offending individuals; in the aggregate, these systems have the potential to change social narratives around communities and to lock in cultural prejudices at scale, replicating and augmenting patterns of structural discrimination and injustice. Likewise, given the scale of their generative abilities, these technologies have broader planetary implications that derive from the cumulative costs of mass industrialized compute. Concentration on the perceived technical complexity of foundation models often mistakenly occludes such a clear ecological view of the social impact of generative outputs.  Regardless, evolving research rightly insists that the AI community must grasp the subtleties, social contexts, and boundaries of human interaction with AI as a user-oriented technology, as well as the social and longitudinal aspects of innovation, more broadly \cite{raji_fallacy_2022, lehmann_mixed-initiative_2023, sanchez_examining_2023}. The risk mapping presented in this paper facilitates this understanding. Even a singular, seemingly innocuous creation of politically incorrect content by an AI tool, a feature that many such tools still permit \cite{hagendorff_ethical_2023}, can accumulate and result in a proliferation of societally effective bias within the community. This bias is amplified by the repetitive contributions from multiple individuals. Consequently, the framework presented here helps us understand how concerns usually detected or faced at individual levels, in fact, scaffold larger risks and contribute to higher-level concerns, thereby providing a deeper socio-technical understanding of foundation models and AI at large.   

\subsection{Toward an integrative perspective on risks and harms}

Our framework illustrates concretely how claims that there may be gains from the implementation and use of foundation models that outweigh their potential and observed harms need to be examined and nuanced through sharpened socio-technical lenses. Such claims often rely on a strictly utilitarian calculation in which the overall potential “benefits” of foundation model application may outweigh the total harms and risks and are hence largely incapable of accounting for more systemic considerations. However, as shown by our framework, foundation model harms are frequently difficult to track and measure—and this affects the capacity to coherently weigh benefits and harms. For instance, harmful use cases that are relatively intelligible when measured individually could have impacts and consequences that are much harder to trace at social and biospheric levels. A moderate, but cumulative and difficult-to-perceive, harm to planetary health has implications for billions of people whereas a significant, but straightforward, harm to some individuals may be quite limited in scope. This makes utility calculations of benefits and harms difficult to perform with sufficient precision to capture their full range, scale, and scope. For example, Bommasani et al. \cite{bommasani_opportunities_2022} argue that the benefits of releasing large models---such as applications that translate text in otherwise underserved languages---outweigh the risks of misuse and abuse by malicious actors. Within their analysis, there is also a recognition that relatively few firms and organizations have sufficient resources and capacities to produce foundation models and that efforts to develop them for use by less-resourced non-elites is nascent and unlikely to produce models with similar capacities. The analysis, however, fails to perceive the deeper risks engendered by the asymmetrical power structures and dynamics that lead to these inequitable differentials in resources, capacities, and access. 

Where a handful of highly self-interested and profit-driven companies control the data, compute, and skills infrastructures on which the development and use of foundation models rely, social harms arising from expanding inequality, wealth polarization, concentration of economic power, and privilege biases that lead to the escalating marginalization of minoritized groups will likely abound. Risk analyses that fail to acknowledge macro-scale issues like this will discount socially consequential adverse impacts. In this case, the very fact that the architecture of foundation models relies on high resource concentration should be a launching pad for expanding the narrow lens on technical risk to account for the ways in which foundation models play a role in reproducing harmful social hierarchies and planetary degradation. Such recognition is also crucial for contending with the potential for foundation models to lead to hard-to-reverse and long-term effects as this technology progressively becomes more embedded in society. What we call an integrative perspective is essential for overcoming the dominant utilitarian and performance-oriented approach in the AI governance discourse which has tended to frame societal challenges in terms of quantifiable trade-offs between risks and benefits.

\section{Conclusion}

In this paper we have confronted a blind spot in the evolving AI governance landscape that derives from its reliance, particularly among prominent policymakers in the global north, on speculative risks and selective seeing. Drawing on the rapidly growing multidisciplinary body of research on foundation-model-generated risks and harms, we have shown how discerning this through integrative and socio-technically curved lenses better discloses the full spectrum of impacts across individual, social, and biospheric levels. We have argued, in this respect, that there exists a visibility gap between the range of concerns and evidence raised in the critical, empirically anchored literature, and the abstract and mainly hypothetical issues focused on within some of the most influential international AI governance initiatives. Such a gap reflects the convergence of the power dynamics, private interests, and geopolitical priorities that have agenda-setting consequences in the AI governance ecosystem---which is a challenge to coherently grappling with the unprecedented industrialization of large scale data-driven technologies, rapidly transforming veritably every domain of life and communities around the world. While potentially beneficial and transformative, foundation models also pose numerous risks to people, society, and the planet. We have aimed to deepen understandings of this broad range of risks by bridging the technical aspects of foundation models with their socio-technical underpinnings, connecting individual concerns to collective and planetary issues, and doing justice to the multifaceted and differential impacts these models have on affected communities. Our conceptualization of risk, particularly regarding the potential transformations effected by these technologies, demonstrates that the visibility of risks and harms should not be concealed or obscured by speculative concerns about existential threats, hypothetically conceived. Instead, as under the framework presented here, an understanding of risks should be grounded in robust evidence that is observable at various levels, demonstrating the potential of these adverse impacts to escalate, over time, and to widen in their scale and scope. Ultimately, the framework we have presented enables a comprehensive assessment of algorithmic impacts for which an interdisciplinary dialogue is key. As such it can be applied as an analytical tool to inform socio-technical mitigations and to fundamentally expand existing toolkits for algorithmic fairness, transparency, and responsible AI.  

\section*{Impact Statement}
1) \textit{Description of the ethical concerns the authors mitigated while conducting the work (as part of an ethical considerations statement):}\newline
Throughout the process of developing the research questions and methodology, we foregrounded our work in an acute awareness and recognition of the role power plays in shaping conversation around the design, development, and deployment of technology. We grounded the evaluation of risks and harms as they relate to foundation models within well-established scholarship and critical discussion surrounding how power is distributed, the pervasiveness of stark power asymmetries, and how the differential impacts of foundation models –and technology at large—are experienced across multiple actors and layers of the ecosystem. We allowed this grounding to inform the framing of our research questions and our methodology, ensuring that a multidisciplinary body of scholarship was drawn on to inform our investigation. This approach included drawing on critical data studies, science and technology studies, and environmental justice scholarship, amongst other fields. It is through this grounding that we aimed to challenge dominant discourses that adopt speculative perspectives on the risks and harms presented by the rapid deployment of foundation models; discourses which overwhelmingly distract from current, real-world consequences. We therefore aim to foreground our work in observed impacts. Through the proposed individual – social – biospheric framework, we work to acknowledge the inherently entangled and interdependent nature of socio-technical systems and their impacts, rather than further dominant discourses of fragmentation and division.

2) \textit{Reflections on how their background and experiences inform or shape the work (as part of a researcher positionality statement):}\newline
The final form and content of this work was shaped by a few factors. Firstly, all researchers involved in the development and writing of this work are academically trained in research, and are based at research and higher-education institutions in the Global North, where the primary language is English. As such, the research itself was limited to the English language. The researchers involved in this work, however, represent a variety of communities and come from a diverse range of backgrounds, including lived and research experience in the Global South. Throughout the process of developing the work, we continuously reflected on-- and engaged with-- how our own values, beliefs, perspectives, and lived experiences inevitably shaped the work presented here. The research team which developed this work is interdisciplinary, coming from a variety of research specializations including anthropology, sociology, data science, and information science, as well as diverse and global industry experience.

3) \textit{Reflection on the adverse, unintended impact the work might have once published (as part of an adverse impact statement):} \newline
In presenting this work, we are acutely aware that the research is squarely grounded in a rapidly evolving field. As such, we acknowledge that the proposed arguments and recommended pathways this research presents might not be relevant in the near to distant future, and we do not prescribe any solutions that could have an impact in the future. However, we also acknowledge that –despite these intentions—our risks and harms codes, and the subsequent categorizations we develop, may have unintended impacts; either through mis-categorizing certain risks and harms, or not accounting for others. These codes and categories were developed by our team’s collective conceptualization of the terms ‘risk’ and ‘harm’; as we are each working within our individual and group positionalities, there may be other forms of risks and harms not fully captured in this research.

\begin{acks}
The authors wish to thank Claudia Fischer and Janis Wong for their key contributions to an earlier draft of this paper, as well as the three anonymous reviewers and area chair for their very constructive feedback. This work was supported by the Ecosystem Leadership Award under the EPSRC Grant EP/X03870X/1, the Arts and Humanities Research Council Grant AH/Z505584/1, and The Alan Turing Institute. 
\end{acks}

%%
%% The next two lines define the bibliography style to be used, and
%% the bibliography file.
\bibliographystyle{ACM-Reference-Format}
\bibliography{references}

%% If your work has an appendix, this is the place to put it.
\newpage
\appendix
\section{Appendix} 
\subsection{Search strategy and objectives}

The primary goal of our search strategy was to identify relevant literature on risks and harms of foundation models and ensure a comprehensive understanding of their socio-political and environmental dimensions. To achieve this objective, we conducted searches across five academic electronic databases. The selected databases include arXiv, ACM Digital Library, IEEE, Scopus, and Web of Science. These databases were chosen based on their relevance in the field, their inclusion of a broader range of sources such as conference proceedings and preprints, and access to the full text of the articles. 

To develop the search strategy, we chose a handful of keywords based on their relevance to the research question and maximize the scope of the search: “foundation model”, “large language model”, “llm”, “general purpose artificial intelligence”, “GPAI”, “risk”, “harm”, “ethic”. In addition, we used Boolean operators to refine search queries (see Table \ref{broad-search} for example). In some cases, we refined when the result set had a high percentage of irrelevant returns: AND TI “artificial intelligence” OR TI “AI” OR AB “artificial intelligence” OR AB “AI”.

Searches were conducted on title and abstract fields. A sample set was piloted, before a full search was conducted across the selected databases. Inclusion criteria covered English language journal articles, including conference papers and pre-prints, but excluding grey literature, monographs, commentaries, correspondences, and opinion pieces. 

The results were added to a Zotero library, which also included our snowball sample, and the merge of these two libraries was exported into a spreadsheet for refinement and coding. We then removed duplicates and non-relevant papers, reaching 167 papers in our sample. The use of an abductive approach supported establishing clearer links with the research objectives and developing a codified mapping of key relationships found in the literature results. To code the papers, firstly, we extracted applicable keyword information from the abstract of each primary study as part of an initial coding process. The resulting keywords served largely as summative and process attributes of a risk or type of harm relevant to foundation models or LLMs. Three researchers subsequently clustered these attributes (Level 2 – attribute codes) and assigned them into Level 1 or parent categories. For example, to attribute code \textit{copyright and intellectual property violations} we assigned the parent category \textit{legal and regulatory violations}. For the attribute codes \textit{difficult to ensure explainability} and \textit{reliance on proprietary software} we assigned the parent category \textit{lock-in and opacity risks} (See Table \ref{tab:parent-attribute-codes} for full list of parent and attribute codes). The codes were subsequently refined by all co-authors.

\begin{table}[H]
\centering
    \setlength{\tabcolsep}{5pt} % Adjust column spacing
\caption{Example of broad search (with refining keywords):}
\begin{tabular}{p{0.45\linewidth} p{0.45\linewidth}}
\hline
\textbf{Broad search (*)} & \textbf{Refining \#2} \\\hline
TI “foundation model*” OR AB “foundation model*” & AND TI “Risk*” OR TI “Harm*” OR TI “ethic*”OR AB “Risk*” OR AB “Harm*” OR AB “ethic*” \\ \hline
TI “Large Language model*” OR TI “LLM*” OR AB “Large Language model*” OR AB ”LLM*” & AND TI “Risk*” OR TI “Harm*” OR TI “ethic*” OR AB “Risk*” OR AB “Harm*” OR AB “ethic*” \\ \hline
TI "general purpose artificial intelligence” OR TI "general-
purpose artificial intelligence” OR TI “GPAI” OR AB “general
purpose artificial intelligence” OR AB "general-purpose
artificial intelligence” OR AB “GPAI” & AND TI “Risk*” OR TI “Harm*” OR TI “ethic*”
OR AB “Risk*” OR AB “Harm*” OR AB “ethic*” \\
\hline
\end{tabular}
\label{broad-search}
\end{table}

\begin{table*}
\centering
  \caption{Parent and attribute codes.}
  \label{tab:parent-attribute-codes}
 \small % Smaller font size
    \setlength{\tabcolsep}{5pt} % Adjust column spacing
\begin{tabular}{p{8cm} p{8cm}}
\hline 
\textbf{Parent Code} & \textbf{Attribute Code}\\ 
\hline
Bias and societal prejudices & Amplifying and perpetuating stereotypes and societal biases \newline Behavioral biases \newline Political biases \\ \hline
Misinformation, disinformation and propaganda & Abusive interactive experiences \newline Manipulation\newline Misinformation spread\\ \hline
Unreliable performance & Accuracy/inaccuracy (outputs) \newline Harmful or toxic outputs  \newline Language performance gap \newline Poor performance due to excessive training with synthetic data \newline Unpredictability of behaviour pre- and post-deployment \newline Untruthfulness of outputs \newline Disparate performance\\ \hline
Cybersecurity risks and harms  & Cyber-attacks payload \newline Data leakage \newline Data poisoning attacks \newline Fraudulent services \newline Goal hijacking and prompt leaking \newline Prompt injection attacks \newline Spear phishing  \newline Jailbreaking \newline Malicious actors \newline Impersonation attack \\ \hline
Privacy risks and harms & Privacy risks and harms \\ \hline
Systemic social and economic risks and harms & Erosion of semantic capital \newline Misleading narratives about AI \newline Narrowing of the market/Monopolisation \newline Outcome homogenization \newline Perpetuation of inequalities \newline Effects on labor market \newline Widening of digital divide \newline Concentration of authoritative power \newline Invisibilization and poor working condition of data and content moderation labor \\ \hline
Legal and regulatory violations & Copyright and intellectual property violations \newline Data protection violations \newline Consumer protection laws violations \\ \hline
Environmental effects and ecological disruption & Environmental effects and ecological disruption \\ \hline
Misuse & Misuse\newline Biological misuse \newline Dual use \newline Illegitimate surveillance \newline Creation of violent or harmful content \\ \hline
Lock-in and opacity risks & Challenges to benchmarking \newline Difficult to ensure explainability \newline Lack of replicability and transparency \newline Low technological readiness \newline Reliance on proprietary software  \\ \hline
Overdependency in human-computer interaction  & Overdependency in human-computer interaction  \\  \hline
Data risks and harms & Data extractivism \newline Data quality \newline Datasets containing toxic data \newline Degradation of the digital commons \\  \hline
Value misalignment & Value misalignment \\ \hline
Extreme or catastrophic risks and harms & Extreme risks \newline Catastrophic risks \\ 
\hline \\[-1.8ex] 
\end{tabular} 
\small
\end{table*} 

\onecolumn
%\newpage
\subsection{Results of the coding and codes description}
\begin{table}[h]
\vspace*{-12pt}
  \caption{Papers per risks and harms parent and attribute codes and codes description.}
  \label{tab:code-descriptions}
\begin{center}
 \small % Smaller font size
    \setlength{\tabcolsep}{5pt} % Adjust column spacing
\begin{tabular}{p{5.29cm}  p{5.29cm}  p{5.29cm}}
\hline \\[-1.8ex] 
Risks and Harms Codes (and Papers)& Code Description & Count \\ 
\hline \\[-1.8ex] 
\textbf{Bias and societal prejudices} \newline Amplifying and perpetuating stereotypes and societal biases: \newline
\cite{abid_persistent_2021, ahuja_mega_2023, alemany_methodology_2023, birhane_multimodal_2021, awais_foundational_2023, bender_dangers_2021, bommasani_opportunities_2022, anderljung_frontier_2023, au_yeung_ai_2023, borchers_looking_2022, brewer_envisioning_2023, chan_reclaiming_2023, connor_large_2023, da_silva_could_2021, huang_survey_2023, dacon_evaluating_2022, de_jager_semantic_2023, deas_evaluation_2023, fan_trustworthiness_2023, farina_chatgpt_2023, felkner_winoqueer_2023, gadiraju_i_2023, glocker_risk_2022, guo_detecting_2021, hossain_misgendered_2023, hosseini_fighting_2023, huang_trustgpt_2023, hundt_robots_2022, khowaja_chatgpt_2023, kidd_how_2023, kirk_personalisation_2023, liao_ai_2023, mattas_chatgpt_2023, mhlanga_open_2023, nadeem_stereoset_2020, nozza_pipelines_2022, oneill_amplifying_2023, ovalle_im_2023, pineiro-martin_ethical_2023, qadir_engineering_2022, rauh_characteristics_2022, sebastian_chatgpt_2023, shaikh_second_2022, shelby_sociotechnical_2023, simmons_moral_2023, solaiman_release_2019, solaiman_evaluating_2023, steed_upstream_2022, sun_safety_2023, talat_you_2022, talboy_challenging_2023, tan_assessing_2019, thakur_unveiling_2023, thieme_foundation_2023, van_der_wal_undesirable_2023, venkit_nationality_2023, wang_decodingtrust_2024, weidinger_taxonomy_2022, weidinger_ethical_2021, wojcik_foundation_2022, zhuo_red_2023, zhao_survey_2023, wu_unveiling_2023, jakesch_co-writing_2023} \newline
Behavioral biases: \newline \cite{chen_manager_2023} \newline Political biases:\newline \cite{liu_mitigating_2021, motoki_more_2023, rozado_political_2023, urman_silence_2023} & The perpetuation and amplification of hegemonic views, harmful stereotypes,
societal and behavioral biases.  & 72 \\ \hline
\textbf{Misinformation, disinformation and propaganda} \newline Abusive interactive experiences: \newline 
\cite{bossert_animals_2021, rosenberg_metaverse_2023, shelby_sociotechnical_2023} \newline Disinformation: \newline 
\cite{anderljung_frontier_2023, barnard_self-diagnosis_2023, bian_drop_2023, christ_undetectable_2023, fecher_friend_2023, kidd_how_2023, liao_differentiate_2023, majovsky_artificial_2023, mattas_chatgpt_2023, neupane_impacts_2023,pan_risk_2023, shelby_sociotechnical_2023, solaiman_release_2019, solaiman_evaluating_2023, weidinger_ethical_2021, weidinger_taxonomy_2022, bommasani_opportunities_2022, li_novel_2023} \newline
Manipulation: \newline
\cite{bender_dangers_2021, chan_reclaiming_2023, kirk_personalisation_2023, shelby_sociotechnical_2023, weidinger_ethical_2021, weidinger_taxonomy_2022, shevlane_model_2023}
\newline 
Hallucinated information: \newline
\cite{connor_large_2023, liao_ai_2023, majovsky_artificial_2023, deroy_how_2023, kang_ethics_2023, tian_opportunities_2023}
\newline
Misinformation spread:\newline
\cite{anderljung_frontier_2023, bian_drop_2023, chan_gpt-3_2023, christ_undetectable_2023, farina_chatgpt_2023, fecher_friend_2023, gao_comparing_2023, goldstein_generative_2023, huang_survey_2023, khowaja_chatgpt_2023, kidd_how_2023, kirk_personalisation_2023, liao_differentiate_2023, liao_ai_2023, mattas_chatgpt_2023, neupane_impacts_2023, pineiro-martin_ethical_2023, qadir_engineering_2022, sebastian_chatgpt_2023, shelby_sociotechnical_2023, solaiman_evaluating_2023, sun_safety_2023, rillig_risks_2023, strasser_pitfalls_2023, thapa_chatgpt_2023, weidinger_ethical_2021, weidinger_taxonomy_2022, zhou_ethical_2023, zhou_synthetic_2023, zhuo_red_2023, li_novel_2023, jakesch_human_2023, tang_science_2023} . & Unintentional or intentional efforts to
disseminate false or misleading
information. Related risks and harms
include disinformation, misinformation
spread, extreme manipulation, and
abusive interactive experiences. \newline  & 45
 \\ 
\hline 
\textbf{Cybersecurity risks and harms} \newline Cyber-attacks payload: \newline
\cite{charan_text_2023, weidinger_ethical_2021, weidinger_taxonomy_2022} \newline
Data leakage: \newline
\cite{bender_dangers_2021, derner_beyond_2023, fan_trustworthiness_2023, huang_survey_2023, jagannatha_membership_2021, liao_ai_2023, oneill_amplifying_2023, pineiro-martin_ethical_2023, plant_you_2022, sebastian_chatgpt_2023, shao_quantifying_2023, wang_decodingtrust_2024, weidinger_ethical_2021, weidinger_taxonomy_2022, zhuo_red_2023, zhou_ethical_2023} \newline
Data poisoning attacks: \newline
\cite{fan_trustworthiness_2023, huang_survey_2023, li_multi-step_2023, wu_unveiling_2023} \newline
Fraudulent services: \newline
\cite{derner_beyond_2023, farina_chatgpt_2023, kirk_personalisation_2023, porsdam_mann_autogen_2023, sebastian_chatgpt_2023, strasser_pitfalls_2023, weidinger_ethical_2021, weidinger_taxonomy_2022}\newline
Goal hijacking and prompt leaking: \newline
\cite{perez_ignore_2022} \newline
Prompt injection attacks: \newline 
\cite{greshake_more_2023, liu_prompt_2023, wu_unveiling_2023, zhuo_red_2023} \newline 
Spear phishing: \newline 
\cite{anderljung_frontier_2023, derner_beyond_2023, eggmann_implications_2023, farina_chatgpt_2023, hazell_large_2023, karanjai_targeted_2022, neupane_impacts_2023, sebastian_chatgpt_2023, solaiman_evaluating_2023} \newline 
Jailbreaking: \newline
\cite{rao_tricking_2023, wang_decodingtrust_2024} \newline
Malicious actors: \newline
\cite{awais_foundational_2023, bommasani_opportunities_2022, eggmann_implications_2023, farina_chatgpt_2023, kang_exploiting_2023, weidinger_ethical_2021, weidinger_taxonomy_2022, wu_unveiling_2023, jakesch_human_2023} \newline
Impersonation attack: \newline
\cite{pineiro-martin_ethical_2023, solaiman_evaluating_2023, zhou_ethical_2023} & Unauthorized access and attacks to the
foundation model that exploit
vulnerabilities or compromise its
integrity. Examples of cybersecurity
risks and harms include cyber-attacks
payload, data leakage, data poisoning
attacks, data pollution, fraudulent
services, goal hijacking and prompt
leaking, offensive cyber capabilities,
prompt injection attacks, spear phishing,
jailbreaking, and other attacks by
malicious actors. \newline  & 37
\end{tabular} 
\end{center}
\end{table} 

\begin{table}
  \label{tab:code-descriptions}
\begin{center}
 \small % Smaller font size
    \setlength{\tabcolsep}{5pt} % Adjust column spacing
\begin{tabular}{p{5.29cm}  p{5.29cm}  p{5.29cm}}
\hline \\[-1.8ex] 
Risks and Harms Codes (and Papers)& Code Description & Count \\ 
\hline \\[-1.8ex] 
\textbf{Unreliable performance} \newline Accuracy/inaccuracy (outputs): \newline 
\cite{au_yeung_ai_2023, awais_foundational_2023, farina_chatgpt_2023, fecher_friend_2023, huang_survey_2023, kang_ethics_2023, liao_ai_2023, oneill_amplifying_2023, pineiro-martin_ethical_2023, shelby_sociotechnical_2023, strasser_pitfalls_2023, zhuo_red_2023, alkaissi_artificial_2023, borji_categorical_2023, rudolph_chatgpt_2023, uz_dr_2023, ye_assessing_2023, casper_explore_2023}
\newline Harmful or toxic outputs:\newline
\cite{anderljung_frontier_2023, bender_dangers_2021, da_silva_could_2021, deshpande_toxicity_2023, huang_trustgpt_2023, liao_ai_2023, borji_categorical_2023, ma_understanding_2023, rauh_characteristics_2022, shaikh_second_2022, shelby_sociotechnical_2023, solaiman_evaluating_2023, strasser_pitfalls_2023, sun_safety_2023, zhuo_exploring_2023, zhuo_red_2023, blodgett_risks_2021, dinan_anticipating_2021, gehman_realtoxicityprompts_2020, li_are_2023, qi_visual_2023, triguero_general_2023, vashishtha_performance_2023, weidinger_ethical_2021, weidinger_taxonomy_2022, welbl_challenges_2021, casper_explore_2023}   \newline  Language performance gap: \newline 
\cite{alshahrani_learning_2022, weidinger_ethical_2021, weidinger_taxonomy_2022, verma_overcoming_2022, zhuo_red_2023} \newline Poor performance due to excessive training with synthetic
data: \newline \cite{veselovsky_artificial_2023}\newline Unpredictability of behaviour pre- and post-
deployment: \newline \cite{anderljung_frontier_2023}\newline Untruthfulness of outputs: \newline \cite{borji_categorical_2023, oviedo-trespalacios_risks_2023, rauh_characteristics_2022} \newline 
Disparate performance: \newline \cite{bommasani_opportunities_2022, shelby_sociotechnical_2023, solaiman_evaluating_2023}
& Inconsistent or undesirable performance
exhibited by foundation models. This includes inaccurate or non-factual
outputs, harmful or toxic outputs,
language performance gap, performance
disparities at group levels, and
unpredictaibility of behavior pre- and
post-deployment. \newline  & 44 \\
\hline
\textbf{Privacy risks and harms} \newline \cite{bender_dangers_2021, eggmann_implications_2023, fecher_friend_2023, huang_survey_2023, jagannatha_membership_2021, kandpal_deduplicating_2022, khowaja_chatgpt_2023, kirk_personalisation_2023, liao_ai_2023, li_multi-step_2023, oneill_amplifying_2023, porsdam_mann_autogen_2023, pineiro-martin_ethical_2023, shelby_sociotechnical_2023, solaiman_evaluating_2023, strasser_pitfalls_2023, tian_opportunities_2023, wang_decodingtrust_2024, weidinger_ethical_2021, weidinger_taxonomy_2022, wu_unveiling_2023, zhou_ethical_2023, patsakis_man_2023, peris_privacy_2023, yan_practical_2023, zhang_right_2023, lu_towards_2023, mireshghallah_empirical_2022} & Adverse consequences associated with
the collection, storage, processing, and
use of data beyond the data pipeline that
arise from the increasing capabilities of
foundation models and their potential to
inadvertently expose private or sensitive
data.  & 29
\\ \hline
\textbf{Systemic social and economic risks and harms} \newline Erosion of semantic capital: \newline 
\cite{nannini_voluminous_2023} \newline  Misleading narratives about AI: \newline
\cite{ferri_risk_2023, nannini_voluminous_2023, shanahan_talking_2023}\newline 
Narrowing of the
market/Monopolisation: \newline 
\cite{bommasani_opportunities_2022, luitse_great_2021} \newline
Outcome homogenization: \newline 
\cite{bommasani_opportunities_2022, kirk_personalisation_2023, porsdam_mann_autogen_2023, thieme_foundation_2023} \newline Perpetuation of inequalities: \newline
\cite{bommasani_opportunities_2022, fecher_friend_2023, kirk_personalisation_2023, porsdam_mann_autogen_2023, rillig_risks_2023, shelby_sociotechnical_2023, solaiman_evaluating_2023, weidinger_ethical_2021, weidinger_taxonomy_2022, williams_framing_2022} \newline
Effects on labor market: \newline 
\cite{farina_chatgpt_2023, khowaja_chatgpt_2023, kirk_personalisation_2023, liao_ai_2023, mattas_chatgpt_2023, eloundou_gpts_2023, lund_chatgpt_2023, qadir_engineering_2022, solaiman_evaluating_2023, weidinger_ethical_2021, weidinger_taxonomy_2022, zarifhonarvar_economics_2023} \newline Widening of digital divide: \newline \cite{khowaja_chatgpt_2023, shelby_sociotechnical_2023} \newline Concentration of authoritative power: \newline \cite{khowaja_chatgpt_2023, solaiman_evaluating_2023} \newline Invisibilization and poor working condition of data and
content moderation labor: \newline \cite{solaiman_evaluating_2023} & Risks and harms that arise from the
widespread adoption and impact of
foundation models on societal and
economic structures. These range from
the erosion of semantic capital, outcome
homogenization, and misleading
narratives about AI, to the widening of
digital divide, the narrowing of the
market or monopolization, effects on
labor market, and the perpetuation of
inequalities.  & 24 \\ 
\hline 
\textbf{Legal and regulatory violations} \newline Copyright and intellectual property violations: \newline
\cite{al-kaswan_abuse_2023, bommasani_opportunities_2022, fecher_friend_2023, farina_chatgpt_2023, khowaja_chatgpt_2023, lund_chatgpt_2023, oneill_amplifying_2023, pineiro-martin_ethical_2023, porsdam_mann_autogen_2023, sebastian_chatgpt_2023, strasser_pitfalls_2023, solaiman_evaluating_2023, lee_speculating_2023, henderson_foundation_2023, huang_generative_2023}
\newline Data protection violations: \newline 
\cite{al-kaswan_abuse_2023, bommasani_opportunities_2022, kandpal_deduplicating_2022, khowaja_chatgpt_2023, kirk_personalisation_2023, oneill_amplifying_2023, patsakis_man_2023, peris_privacy_2023, pineiro-martin_ethical_2023, porsdam_mann_autogen_2023, sebastian_chatgpt_2023, solaiman_evaluating_2023, strasser_pitfalls_2023, tian_opportunities_2023, zhang_right_2023} \newline
 Consumer protection laws violations: \newline \cite{sebastian_chatgpt_2023} & Breaches of laws and regulations,
including those designed to safeguard
individuals’ privacy and secure handling
of their personal data, intellectual
property, consumer protection, and
cybersecurity.  & 21
\end{tabular} 
\end{center}
\end{table} 

\begin{table}
  \label{tab:code-descriptions}
\begin{center}
 \small % Smaller font size
    \setlength{\tabcolsep}{5pt} % Adjust column spacing
\begin{tabular}{p{5.29cm}  p{5.29cm}  p{5.29cm}}
\hline \\[-1.8ex] 
Risks and Harms Codes (and Papers)& Code Description & Count \\ 
\hline \\[-1.8ex] 
\textbf{Environmental effects and ecological disruption} \newline 
\cite{anthony_carbontracker_2020, bender_dangers_2021, bommasani_opportunities_2022, hagendorff_speciesist_2023, huang_survey_2023, khowaja_chatgpt_2023, kirk_personalisation_2023, kirkpatrick_carbon_2023, liao_ai_2023, luccioni_estimating_2022, oneill_amplifying_2023, rillig_risks_2023, shelby_sociotechnical_2023, solaiman_evaluating_2023, triguero_general_2023, weidinger_taxonomy_2022, patterson_carbon_2021, li_making_2023} & Adverse and negative impacts of
foundation models on ecology and
animals.  & 19
\\ 
\hline 
\textbf{Misuse} \newline Misuse: \newline 
\cite{anderljung_frontier_2023} \newline Biological misuse: \newline \cite{cohen_what_2023, sandbrink_artificial_2023} \newline Dual use: \newline 
\cite{anderljung_frontier_2023, kang_exploiting_2023, weidinger_ethical_2021, weidinger_taxonomy_2022, glukhov_llm_2023, soice_can_2023, henderson_self-destructing_2023}
\newline  Illegitime surveillance: \newline
\cite{anderljung_frontier_2023, bommasani_opportunities_2022, kirk_personalisation_2023, weidinger_ethical_2021, weidinger_taxonomy_2022} \newline Creation of violent or harmful content: \newline 
\cite{anderljung_frontier_2023, bender_dangers_2021, bommasani_opportunities_2022, kirk_personalisation_2023, shelby_sociotechnical_2023, solaiman_evaluating_2023} & The misuse of foundation models is used
here to describe specific contexts in
which foundation models or
technologies based on foundation
models may be inappropriately used and
lead to negative consequences. This
includes uses that can have both civilian
and military applications, or contribute
to the handling, manipulation, or
application of biological materials,
organisms, or technologies.  & 19 \\
\hline
\textbf{Lock-in and opacity risks} \newline Challenges to benchmarking: \newline
\cite{awais_foundational_2023} \newline Difficult to ensure
explainability: \newline 
\cite{pineiro-martin_ethical_2023, thieme_foundation_2023, triguero_general_2023, wasserman_rozen_case_2023}\newline Lack of replicability and transparency:\newline \cite{farina_chatgpt_2023, hosseini_fighting_2023, kang_ethics_2023, khowaja_chatgpt_2023, liao_ai_2023, oneill_amplifying_2023, pineiro-martin_ethical_2023, yan_practical_2023, zhou_ethical_2023, lu_towards_2023} \newline Low technological readiness: \newline \cite{yan_practical_2023}\newline Reliance on proprietary software: \newline \cite{liao_ai_2023, liesenfeld_opening_2023} & Risks and harms associated with foundation models that are difficult to understand, replicate, or modify. These risks and harms arise from issues including reliance on proprietary software, challenges to benchmarking, lack of transparency, difficulty to ensure explainaibility, and low technological readiness. & 15 \\ 
\hline 
\textbf{Overdependency in human-computer interaction} \newline 
\cite{connor_large_2023, kirk_personalisation_2023, liao_ai_2023, ma_understanding_2023, mhlanga_open_2023, pineiro-martin_ethical_2023, solaiman_evaluating_2023, thieme_foundation_2023, weidinger_ethical_2021, weidinger_taxonomy_2022, peris_privacy_2023}
 & Overreliance of humans
interacting with computer systems,
interfaces, and technologies built on
foundation models for various aspects of
their lives and potentially negatively
impacting their well-being, safety,
decision-making abilities, or
interpersonal relationships. & 11 \\
\hline
\textbf{Data risks and harms} \newline Data extractivism: \newline \cite{huang_generative_2023} \newline Data quality: \newline \cite{huang_generative_2023, kandpal_deduplicating_2022} \newline Datasets containing toxic data: \newline \cite{birhane_science_2023, fan_trustworthiness_2023, oneill_amplifying_2023, pineiro-martin_ethical_2023, wang_decodingtrust_2024} \newline Degradation of the digital commons: \newline \cite{chan_gpt-3_2023, huang_generative_2023} & A range of risks and harms associated
with the collection, storage, processing, and use of data within the data pipeline.
These risks extend beyond the cybersecurity risks and harms described above and includes broader issues of ethical, social, and economic implications of data stewardship and management. They include issues of poor data quality, datasets containing toxic data, data extractivism, and the degradation of the digital commons. & 8
\\ \hline
\textbf{Value misalignment} \newline \cite{huang_generative_2023, kasirzadeh_conversation_2023} & Misalignment of the output content of
conversational agents with the norms
and values of the human interacting with
such agent. & 2 \\ 
\hline 
\textbf{Extreme or catastrophic risks and harms} \newline Extreme risks: \newline \cite{shevlane_model_2023} \newline Catastrophic risks: \newline \cite{barrett_actionable_2023} & Speculative far-reaching or irreversible
adverse impacts of foundation models at
societal scale that extend beyond
immediate impacts. & 2
\end{tabular} 
\end{center}
\end{table} 

\end{document}